\documentclass{jpp}
\usepackage{graphicx}

\usepackage[utf8]{inputenc}
\usepackage[T1]{fontenc}
\usepackage{amsmath}
\usepackage{comment}

\shorttitle{Relativistic tearing instability}
\shortauthor{I. Demidov and Y. Lyubarsky}

\title{Collisionless tearing instability in a relativistic pair plasma with a power-law distribution function}

\author{
  I. Demidov \corresp{\email{dvsmallville@gmail.com}}
  \and
  Y. Lyubarsky}

\affiliation{Physics Department, Ben-Gurion University, PO Box 653, Beer-Sheva 84105, Israel}

\begin{document}

\maketitle

\begin{abstract}
We study the tearing instability of a current sheet in a relativistic pair plasma with a power law distribution function. We first estimate the growth rate analytically and then confirm the analytical results by solving numerically the dispersion equation, taking into account
all exact particle trajectories within the reconnecting layer. We found that the instability is suppressed when the particle spectrum becomes harder. 
\end{abstract}

\section{Introduction}

Collisionless magnetic reconnection plays crucial role in many astrophysical phenomena, such as solar flares \citep{Innes2015}, pulsar winds and nebulae \citep{Coronity1990,Lyubarsky2001},  jets from active galactic nuclei \citep{Romanova1992,Christie2019}, gamma-ray bursts \citep{McKinney2012,Lazarian2019} and black hole magnetospheres \citep{Bransgrove2021,Ripperda2022}. This process provides efficient conversion of the magnetic to the particle energy. 
Magnetic reconnection could effectively produce nonthermal particle distributions 
\citep{Zenitani2001,Larrabee2003,
Drake2013,Guo2020,Uzdensky2022}. 

In this paper, we consider 
the linear stage of tearing instability in a relativistic pair plasma. This instability plays a key role in triggering magnetic reconnection. Even though non-relativistic tearing instability has been studied intensively, only a few papers generalise these results to the relativistic plasma. 
\cite{Zelenyi1979} considered the collisionless tearing mode in the relativistic plasma analytically by using the kinetic approach. Their results were confirmed both analytically and numerically by \cite{Petri2007,Zenitani2007, Hoshino2020}. 
Relativistic tearing instability was analytically studied in the fluid approach by \citet{Lyutikov2003, Komissarov2007, Yang2019}. 

Fluid models applied to a collisionless pair plasma can take into account kinetic effects only phenomenologically 
by including inertial terms and off-diagonal pressure terms in the generalised Ohm's law (e.g. \citealt{Hesse2007}). In the simplest geometry without a guide field, these additional terms can be reduced to the form that includes effective or "anomalous" resistivity (\citealt{Bessho2012}) that arises from  kinetic effects of resonant wave-particle interactions (e.g. \citealt{Coppi1966}, \citealt{Galeev1975}).
Additionally, the collisionless plasma can be considered collisional due to strong microscopic turbulence that scatters particles similarly to collisions in a resistive plasma 
(e.g. \citealt{Lyutikov2003}, \citealt{Komissarov2007}, \citealt{Elenbaas2016}); in extreme cases
the collisional state 
can be sustained by frequent
pair annihilation and reconversion, $e^++e^-\leftrightarrow\gamma$ \citep{ThompsonKostenko2020}.  Despite the simplicity of the fluid equations, which do not contain information about complicated particle trajectories and various microscopic processes, kinetic theory is needed to determine the effective resistivity, which is usually considered as a free parameter of the problem and, therefore, makes the fluid approach incomplete. 

The thermal plasma with constant temperature was considered in the above kinetic studies of collisionless tearing instability. 
However, the plasma in relativistic astrophysical sources could hardly reach the Maxwellian distribution since binary particle collisions are typically too rare due to low plasma density and high temperatures. Therefore, one has to consider the class of non-thermal distribution functions with a wide energy spread, which is relevant for such conditions.
We consider the simplest case without the guide magnetic field, when the distribution function has the power-law form, 
\begin{equation}\label{pwl}    
\text{d}\mathcal{N}(\mathcal{E})=C\mathcal{E}^{-\alpha}\exp\left(-\frac{\mathcal{E}}{\mathcal{E}_\text{max}}\right)\text{d}\mathcal{E}, \quad \text{at} \,\,\, \mathcal{E}\geq\mathcal{E}_\text{min}
\end{equation}
where $C$ is a normalisation constant, $\alpha$ is a spectral index and $\mathcal{E}_\text{max}$ is the cut-off energy. At energies $\mathcal{E}<\mathcal{E}_\text{min}$ we assume that $\text{d}\mathcal{N}/\text{d}\mathcal{E}=\text{const}$.

Collisionless tearing instability in the relativistic pair plasma with the non-Maxwellian distribution function of particles was recently studied by \cite{Thompson2022}. He considered specific conditions in the pulsar emission zone, namely, weakly sheared quantizing magnetic field and  the narrow top hat distribution function centered at characteristic momentum $p_0$ both for electrons and positrons. 

Our aim is  to study how the presence of the high-energy tail in the particle spectrum affects the tearing mode. We obtain the growth rate of the tearing instability in the linear regime and find how it depends on the power-law spectral index, $\alpha$.

The article is organized as follows: In Section 2, we obtain analytical estimates for the growth rate of the tearing instability. In Section 3, a numerical calculation of the growth rate is carried out 
to confirm the obtained analytical estimates. In Section 4, we summarize our results.

\section{Analytical estimates}

\subsection{Preliminary considerations}

\begin{figure}
  \centering
  \includegraphics[width=\columnwidth]{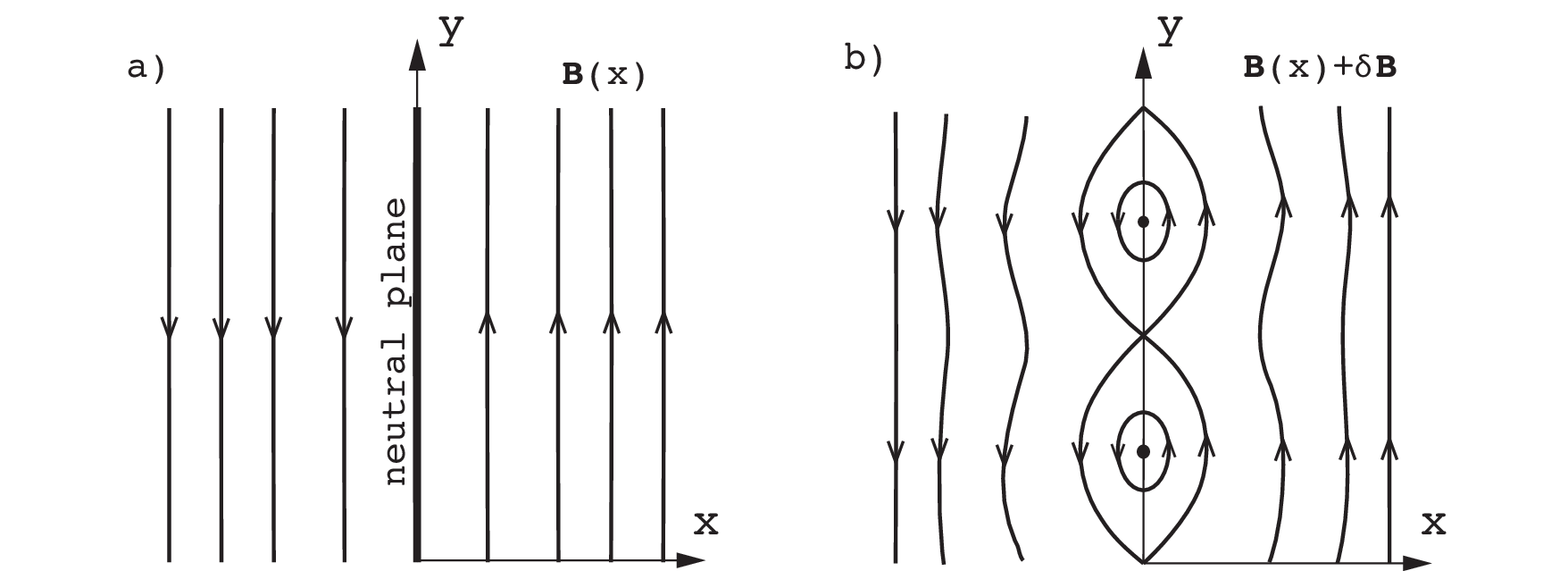}
  \caption{Magnetic field configuration within the current sheet: a) unperturbed; b) with the tearing mode perturbation}
\label{f0}
\end{figure}

Let the unperturbed reconnecting magnetic field be directed along $y$-axis $\boldsymbol{B}=B_0f(x)\hat{\boldsymbol{y}}$, where $f(x)$ is a magnetic field profile that depends only on $x$-coordinate (see Figure~\ref{f0}a). In such a geometry, the magnetic field is presented by the vector potential $\boldsymbol{A}=(0,0,A_z(x))$. In equilibrium, there is no electric field, and electrons and positrons drift in the $z$-direction at a velocity $\pm U$ ("$+$" for positrons and "$-$" for electrons, $U>0$). 
The pair plasma under consideration is ultrarelativistic; therefore, the approximate dispersion law for plasma particles $\mathcal{E}\approx pc$ is used.

We assume that the current sheet thickness is much greater than the Larmor radius of the particles; therefore, with the exception of the neutral layer at small $x$, the MHD approximation is valid throughout the entire space, regardless of the form of the distribution function. Namely, the static equilibrium is determined by the balance of the magnetic pressure and the effective particle pressure found as the appropriate moment of the distribution function \citep[pp. 217-218]{Alfven1963}. 
One can use a prepared magnetic field profile, and it is assumed that the distribution function of plasma is adjusted to this field. For simplicity, we consider the current sheet with the magnetic field profile $f(x)=\tanh(x/L)$ \citep{Harris1962}. Near the neutral layer, the MHD approximation is violated, and kinetic effects are important. Thus, one can find solutions in the inner and outer regions and then match them together, which ultimately gives us an expression for the growth rate of the tearing instability.




The general picture of collisionless tearing instability has long been known (e.g. 
\citealt[pp. 309-315]{Galeev1984}, \citealt{Kadomtsev1987}).
Let us consider the equilibrium of the current sheet as a whole. 
The plasma pressure near the neutral plane $x=0$ is balanced with the magnetic pressure at infinity, i.e. $P_\text{tot}(0)=B_0^2/8\pi $.   The plasma pressure can be written as
$ P_\text{tot}(0)=(2/3)n_0\langle \mathcal{E}\rangle$,
where $\langle \mathcal{E}\rangle$ is the average energy of particles. We also can consider the same balance but in terms of forces, i.e. $\nabla P_\text{tot}=(1/c)\,\boldsymbol{j}\times\boldsymbol{B}$. Roughly, one can estimate $\nabla P_\text{tot}\sim P_\text{tot}(0)/L$, where $L$ is the characteristic thickness of the considered layer and the total current density can be estimated as $j_z\sim 2en_0 U$. As a result, we obtain the following two equations that describe the current sheet equilibrium
\begin{equation}\label{two}
    B_0^2=\frac{16\pi}{3} n_0\langle\mathcal{E}\rangle, \quad \frac{U}{c}\sim \frac{\langle\mathcal{E}\rangle}{eB_0 L}.
\end{equation}
The second equation is just a definition of the characteristic diamagnetic drift velocity of the plasma near the neutral plane (see Appendix~\ref{appA} for additional details). It is important to note that high-energy particles play a significant role only at $\alpha<2$ since in this case, the average energy is $\langle\mathcal{E}\rangle\sim \mathcal{E}_\text{max}$, and at $\alpha\geq 2$ we have $\langle\mathcal{E}\rangle\sim\mathcal{E}_\text{min}$ (i.e. the high-energy tail is suppressed).

The current sheet that is described above can be considered as a set of straight filaments with a current. Since the currents in filaments flow in the same direction, they are attracted to each other, so such an equilibrium system with a current sheet is unstable to pinching \citep{Coppi1966}. In fact, this is possible only if the resistivity of plasma is not zero.  
Even small non-zero resistivity allows the currents to move, and any small displacement of a filament gives rise to a force imbalance. As a result, the individual current filaments merge, leading to a reconnection of the magnetic field and the formation of magnetic islands (see Figure~\ref{f0}b).

Effective resistivity arises in the collisionless plasma because the particles are not magnetised near the plane $x=0$, so they can gain energy in the electric field. 
The periodic electric field, $\delta \boldsymbol{E}$, is induced near the neutral plane $x=0$ due to the magnetic reconnection. Indeed, according to Lenz's law, 
the electric field is directed in the negative $z$-direction in the vicinity of X-points, where plasma flows towards the neutral plane, increasing the magnetic flux, and in the positive $z$-direction between the  X-points, where plasma flows away from the neutral plane. 
In turn, the particles are not magnetised in the layer with the characteristic thickness \citep{Parker1957},
\begin{equation}
l\sim \sqrt{2r_{g0} L},
\end{equation}
where $r_{g0}=p_\perp c/eB_0$ is the particle Larmor radius in the asymptotic magnetic field. 
Inside this layer, particles perform a meandering motion along the current sheet 
and, therefore, they are accelerated by the induced electric field. 
The acceleration time is $\sim 1/kc$, 
because at a larger time, when the particle moves along the $y$-axis, it "sees" an oscillating field and does not gain energy on average. 

Thus, one can estimate the effective conductivity within the inner region by using the Drude formula with the effective collision time $\tau_\text{coll}\sim 1/kc$ \citep[p. 242]{Artsimovich1979},
which gives 
\begin{equation}\label{conduct}
    \sigma^\text{eff}\sim \frac{2n_0 e^2}{m}\frac{1}{k c}\bigg\langle\frac{mc^2}{ \mathcal{E}}\bigg\rangle.
\end{equation}
Here, angular brackets denote averaging over particle energies.
The factor $\langle mc^2/\mathcal{E}\rangle$ is just the average inverse Lorentz factor of particles; it reflects the fact that conductivity in relativistic plasma is inversely proportional to the average relativistic mass $m_{\rm eff}=\mathcal{E}/c^2$. Since the averaged energy of the particles is in the denominator, the non-magnetised particles with the lowest energy make the largest contribution to the conductivity. 
According to this, we can assume that $\sigma^\text{eff}$ operates only within the "inner" region of the size 
\begin{equation}
l_\text{min}\sim \sqrt{\frac{2L}{eB_0\langle\mathcal{E}^{-1} \rangle}}.
\end{equation}
Let us emphasize that in our case, the flux-freezing condition is violated because there is no magnetic field near the neutral layer, and the particles there are not magnetised such that they decouple from the fluid motion. This is what determines the thickness $l$ of the inner layer where the particles can be accelerated by the electric field. Here, the effective resistivity appears simply because the particles have inertia and "lazily" respond to the electric field when moving within the inner region during the "collisional" time $1/kc$.



\subsection{The outer region}

Before estimating the growth rate of tearing instability, let us consider the outer region $|x|>l_\text{min}
$, where the plasma can be considered as an ideal magnetised fluid with infinite conductivity $\sigma\rightarrow\infty$. Indeed, since the Larmor radius of the particles here is much smaller than the characteristic thickness of the current sheet $L$, one can use the ideal MHD approximation. Even though local thermodynamic equilibrium is not valid in the considered problem, and strictly speaking, we cannot use MHD equations, tearing instability develops slowly, so at each moment of time, we can consider the problem as quasi-static and consider the plasma velocity as a small perturbation. The validity of this statement is also confirmed by the fact that the kinetic theory gives the same equations (see Appendix A). 

\begin{figure}
  \centering
  \includegraphics[width=\columnwidth]{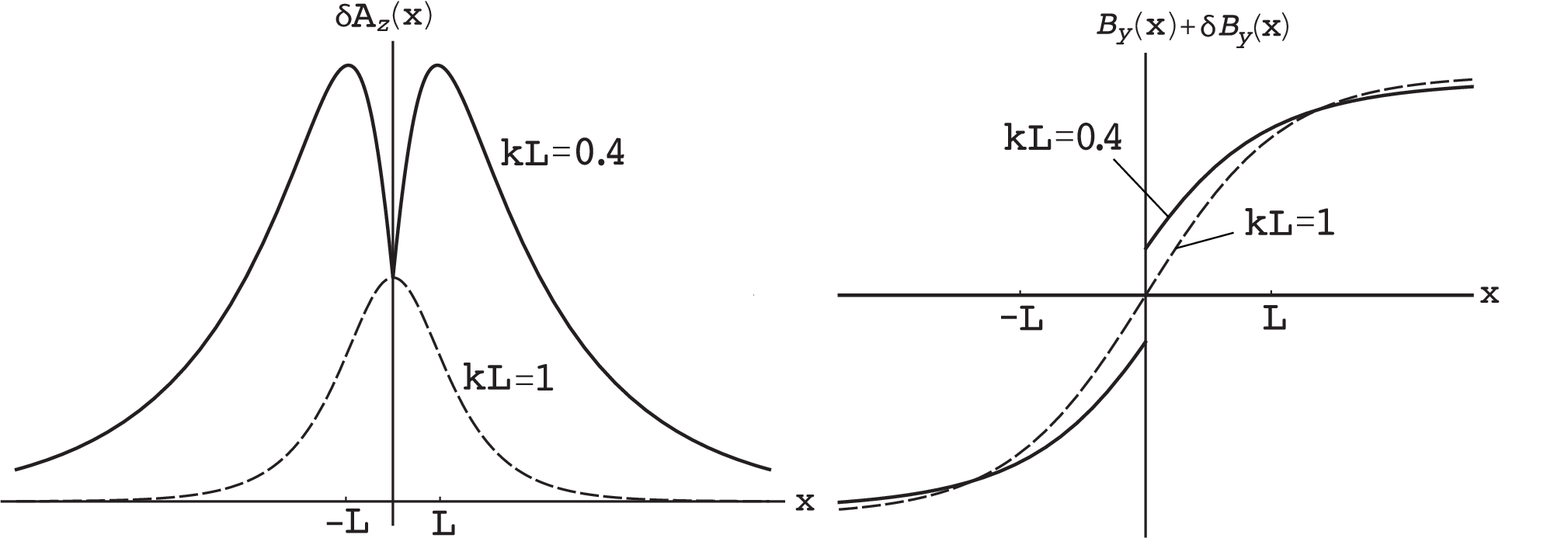}
  \caption{The normalized vector potential perturbation and the total magnetic field for the Harris current sheet at $kL=0.4$ and $kL=1$}
\label{Harris}
\end{figure}

Slow plasma motions create the current density perturbation, which, in turn, creates a disturbance of the magnetic field, such that the total magnetic field can be represented as $\boldsymbol{B}(x,y)=B_0f(x)\hat{\boldsymbol{y}}+\delta\boldsymbol{B}(x,y)$. The magnetic field perturbation $\delta\boldsymbol{B}$ can be expressed via the vector potential perturbation $\delta\boldsymbol{A}=\delta A_z\hat{\boldsymbol{z}}$ according to $\delta\boldsymbol{B}=\text{rot}\,\delta \boldsymbol{A}$. Considering the magnetic field profile $f(x)=\tanh(x/L)$, Ampère's equation is reduced to (e.g. \citealt{Sturrock1994}, \citealt{Boldyrev2018})
\begin{equation}
\frac{\text{d}^2}{\text{d}x^2}\delta A_z+\left[-k_y^2+\frac{2}{L^2\cosh^2(x/L)}\right]\delta A_z=0,
\end{equation}
Therefore, the even-parity vector potential perturbation in the outer layer is (\citealt{Sturrock1994}, \citealt{Boldyrev2018})
\begin{equation}\label{outer}
\delta A_z(x)=\delta A_z(0)\left(1+\frac{1}{kL}\tanh\frac{|x|}{L}\right)\exp\left(-k|x|\right),
\end{equation}
where $k=|k_y|$. Odd-parity solutions $\delta A_z(-x)=-\delta A_z(x)$ are responsible for kink modes and not interesting to us. It is clearly seen from the solution~(\ref{outer}) that the vector potential perturbation is continuous at $x=0$ for all $kL$, but the derivative $\delta A'_z(x)$ is discontinuous in the general case. Let us introduce the notation
\begin{equation}\label{delta}
    \Delta'(0)=\frac{\delta A'_z(0+)-\delta A'_z(0-)}{\delta A_z(0)}\equiv\frac{\delta B_y(0+)-\delta B_y(0-)}{\text{i}(c/\omega)\delta E_z(0)}.
\end{equation}
According to~(\ref{outer}), we obtain
\begin{equation}\label{Hdelta}
    \Delta'(0)=\frac{2(1-k^2L^2)}{kL^2}.
\end{equation}
Therefore, $\Delta'(0)=0$ and there is no discontinuity $[\![\delta B_y]\!]=0$ if and only if $kL= 1$ (see Figure~\ref{Harris}). 
At $kL\neq 1$ the magnetic field perturbation jump occurs. To remove the discontinuity of the magnetic field, we have to take into account the current of non-magnetised particles in the region $|x|<l_\text{min}
$. 

\subsection{Tearing instability growth rate}

Let us denote the current density of particles undergoing meandering motion near the neutral plane as $\delta j_z$. We assume for simplicity that $\delta j_z$ vanishes at $|x|>l_\text{min}$. According to Ampère's law, the magnetic field jump $[\![\delta B_y]\!]=\delta B_y(0+)-\delta B_y(0-)$ can be found as
\begin{equation}\label{amp}
     [\![\delta B_y]\!]\approx \frac{4\pi}{c}\!\!\int\limits_{-l_\text{min}}^{l_\text{min}}\!\!\delta j_z\text{d}x.
\end{equation}
Therefore, $\delta j_z$ eliminates discontinuity and provides a smooth transition of the perturbed magnetic field accross the plane $x=0$.

According to Ohm's law, $\delta j_z=\sigma^\text{eff}\delta E_z$. Therefore, the current density of non-magnetised particles is directed along the electric field $\delta E_z$. On the other hand, the sign of the magnetic field jump is determined by the outer solution, and according to~(\ref{delta}), one can write $[\![\delta B_y]\!]=(c/\gamma)\Delta'(0)\delta E_z(0)$, where $\gamma=-\text{i}\omega$ is the growth rate of the instability. At $kL>1$, the direction of the current coincides with the sign of the magnetic field jump only if $\gamma<0$ and the instability is suppressed. Therefore, the tearing mode is arising unstable only for $kL<1$.

Further, it is convenient to 
express the electromagnetic fields via the vector potential perturbation $\delta A_z$: 
\begin{equation}
    \delta E_z=\frac{\text{i}\omega}{c}\delta A_z\equiv -\frac{\gamma}{c}\delta A_z, \quad \delta B_y=-\frac{\text{d}}{\text{d}x}\delta A_z.
\end{equation}
Assuming that the electric field is approximately constant within the inner region of size $\sim 2l_\text{min}$, we obtain from~(\ref{amp}) that
\begin{equation}\label{incr}
    \gamma\approx \frac{c^2}{8\pi}\frac{\Delta'(0)}{\sigma^\text{eff} l_\text{min}}.
\end{equation}
It should be noted that this result can be used for different magnetic field profiles. One has to calculate $\Delta'(0)$ for the given profile and substitute it into the above equation.  

Substituting the effective conductivity~(\ref{conduct}) and expression for $\Delta'(0)$ into equation~(\ref{incr}) and using relations for the current sheet equilibrium~(\ref{two}) we obtain
\begin{equation}\label{grwth}
    \gamma(k)\sim\frac{c}{2L}(1-k^2L^2)\left(\frac{U}{c}\right)^{3/2}\left[\langle \mathcal{E}\rangle \langle \mathcal{E}^{-1}\rangle\right]^{-1/2}
\end{equation}
In general case, characteristic energies $\langle \mathcal{E}\rangle$ and $1/\langle \mathcal{E}^{-1}\rangle$ may not coincide. At large $\alpha\geq 2$, we expect $\langle \mathcal{E}\rangle\sim\langle\mathcal{E}^{-1}\rangle^{-1}\sim\mathcal{E}_\text{min}$ and the growth rates are determined by the same formula as for the ultrarelativistic Maxwellian plasma \citep{Zelenyi1979}. On the other hand, in the case of the power-law distribution with $\alpha<2$, one can expect that $\langle \mathcal{E}\rangle\sim \mathcal{E}_\text{max}$ and $\langle \mathcal{E}^{-1}\rangle^{-1}\sim \mathcal{E}_\text{min}$, i.e., there are two characteristic energy scales. Then, the growth rate contains a small factor $\epsilon^{1/2}$, where
\begin{equation}
\epsilon=\frac{\mathcal{E}_\text {min}}{\mathcal{E}_\text{max}}.
\end{equation}

According to equation~(\ref{grwth}), the growth rate reaches a non-zero constant value when $k$ goes to zero. However, since the conductivity $\sigma^\text{eff}\sim 1/k$ decreases with $k$, the effective resistance disappears, and there should be $\gamma\rightarrow 0$ in the limit $k\rightarrow0$. 
Therefore, equation~(\ref{grwth}) is invalid at small $k$. Indeed, in this case, we cannot assume that the electric field is constant within the region of size $\sim l_\text{min}$, such that $\delta E_z(l_\text{min})\approx \delta E_z(0)$. 

To find the correct asymptotic, one has to solve the Maxwell equations within the inner region. By using zero-divergence of the magnetic field and Faraday's law,
\begin{equation}
    \frac{\partial\delta B_x}{\partial x}+\text{i}k \delta B_y=0, \quad \delta B_x=\frac{ck}{\omega}\delta E_z,
\end{equation}
one can rewrite Ampère's law in terms of the electric field perturbation only
\begin{equation}
    \frac{\text{d}^2}{\text{d}x^2}\delta E_z-\frac{4\pi\gamma}{c^2}\sigma^\text{eff}\delta E_z=0,
\end{equation}
where the limit $k\rightarrow 0$ and the boundary condition $\text{d}(\delta E_z)/\text{d}x=0$ at $x=0$ are assumed. Thus, the electric field perturbation profile within the inner region can be written as
\begin{equation}
    \delta E^{(\text{i})}_z(x)\approx \delta E_z^{(\text{e})}(l)\frac{\cosh \eta x}{\cosh \eta l},
\end{equation}
where $\eta=\sqrt{4\pi\gamma\sigma^\text{eff}/c^2}$ and $l=l_\text{min}$. 
Performing integration in~(\ref{amp}), we obtain the dispersion equation at $k\rightarrow0$:
\begin{equation}
\Delta'(l)=2\eta\tanh\eta l,
\end{equation}
where now the tearing parameter is
\begin{equation}\label{parameter1}
    \Delta'(l)=\frac{\delta B_y(l)-\delta B_y(-l)}{\text{i}(c/\omega)\delta E_z(l)}.
\end{equation}
Taking into account the outer solution~(\ref{outer}), in the limit $k\rightarrow 0$ we obtain
\begin{equation}
    \delta A_z(x)\approx \frac{\delta A_z(0)}{kL}\tanh\frac{|x|}{L}, \quad \Delta'(l)\approx \frac{2}{L}\frac{\text{sech}^2(l/L)}{\tanh(l/L)}\approx\frac{2}{l}.
\end{equation}
This leads to the following dispersion equation
\begin{equation}\label{est}
\eta l\tanh \eta l\approx 1.
\end{equation}
In order of magnitude, this equality is satisfied when the tangent argument is of the order of unity, therefore
\begin{equation}\label{est1}
    \gamma(k)\sim\frac{c^2}{4\pi}\frac{1}{\sigma^\text{eff}(l_\text{min})^2}\sim \frac{1}{3}ck\!\left(\frac{U}{c}\right).
\end{equation}
This expression gives the correct asymptotic behavior for the growth rate of instability at $k\rightarrow0$. 



Obviously, different magnetic profiles lead to different scalings of the corresponding tearing instability growth rates. In the context of MHD turbulence, a periodic magnetic field $B(x)\sim B_0\sin(x/L)$ is more appropriate \citep{Boldyrev2018}. However, dependence on $\alpha$ and $\mathcal{E}_\text{min}/\mathcal{E}_\text{max}$ is unchanged for any magnetic field profile with the same asymptotic behavior at $x\rightarrow 0$. Therefore, in this work, we focus only on the usual Harris layer.

\section{Numerical Solution}

\begin{figure}
  \centering
  \includegraphics[width=\columnwidth]{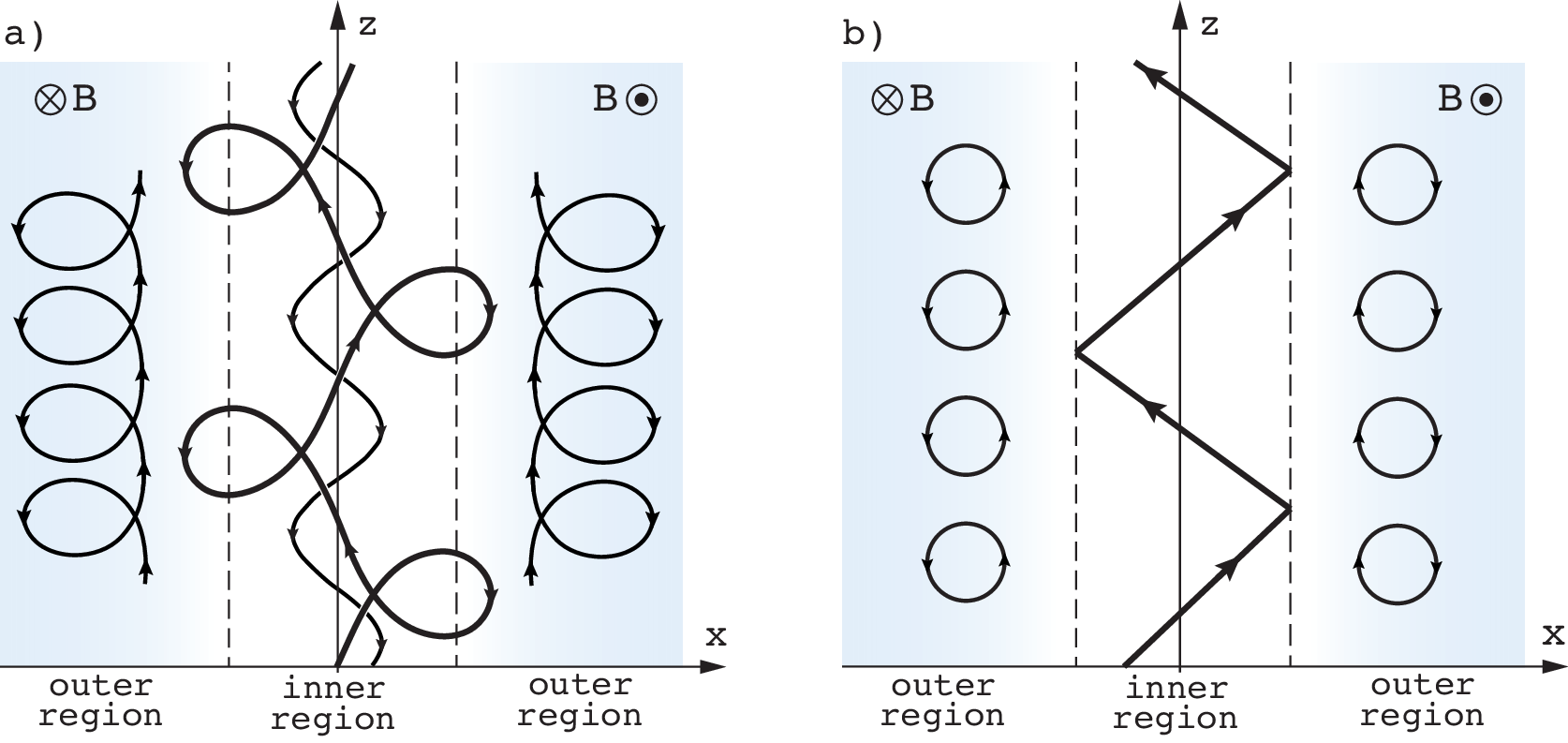}
  \caption{Schematic representation of unperturbed particle orbits: a) exact orbits; b) approximate orbits}
\label{f1}
\end{figure}

In this section, we numerically solve the equation for the perturbed vector potential and determine the tearing instability growth rate, comparing the result with our qualitative estimations. According to the well-known procedure (see Appendix~\ref{appA}), the Ampère equation for the vector potential perturbation can be written in the form
\begin{equation}\label{v}
\begin{split}
\frac{\text{d}^2}{\text{d}x^2}\delta A_z+\Big[-k_y^2+&\frac{2}{L^2\cosh^2(x/L)}\Big]\delta A_z= \\
&=4\pi\text{i}\omega \sum_s\frac{q_s^2}{c^2}\int \frac{\partial f_{0s}}{\partial \mathcal{E}}v_z\int\limits_{-\infty}^t v_z(t')\delta A_z(t')\text{d}t' \text{d}\boldsymbol{p}.
\end{split}
\end{equation}
The main difficulty lies in the R.H.S. of the equation, which represents the current of non-magnetised particles. Here it is necessary to integrate along rather complex particle orbits, which are shown schematically in Figure~\ref{f1}a. These orbits are solutions of the unperturbed equations of motion, i.e., only when the stationary magnetic field $B(x)=B_0\tanh(x/L)$ is present.

In early works, the neutral sheet was divided into two regions: inner and outer. Together with the assumption of constancy of the vector potential perturbation $\delta A_z$ within the internal region, this allows the integro-differential equation~(\ref{v}) to be reduced to the ordinary "Schrödinger-type" differential equation. In the inner region $|x|\lesssim  l\sim \sqrt{2r_{0g}L}$, magnetic field is weak, and this region is dominated by particles whose orbits cross the resonance plane $x=0$. 
\citet{Laval1966} and \citet{Hoh1966} refer to the complicated actual particles orbits within the inner region. \citet{Coppi1966} proposed a simplified model of the particle orbits, which are approximated by straight line segments within the inner region and Larmor circles in the outer one (see Figure~\ref{f1}b). By using such an orbit model, \citet{Dobrowolny1968} gave a quantitative calculation of the linear growth rate of tearing instability. By comparing his results with \citet{Laval1966}, \citet{Hoh1966}, he claimed that the complexity of the actual particles trajectories within the current sheet is not important in the instability mechanism.
Therefore, these orbits are often taken to be straight lines along the neutral plane (e.g. \citealt{Galeev1975}). This approach gives the same answer as piece-wise straight orbits since particles execute rapid oscillations between magnetic "walls" $x=\pm l$ with frequency $\sim v_T/l$, and the motion averaged over these oscillations can be considered as free motion along straight lines. The outer region $|x|>l$  is dominated by "non-crossing" particles, which are assumed to have small Larmor orbits and these orbits are neglected. Indeed, since in the limit $r_g\rightarrow 0$, the magnetic field changes slightly along a Larmor radius of non-crossing particles; therefore, orbits are not strongly distorted and the drift of the guiding centre is slow. \citet{Zelenyi1979} solved the considered problem for relativistic Maxwellian electron-positron plasma in the same way.

We solve the problem numerically, taking into account all unperturbed particle orbits in the integro-differential equation~(\ref{v}) and removing two restrictive simplifications: "constant-$\delta A$" approximation at $|x|<l$ and the use of approximate orbits. 
Approaches for the numerical solving of the equation~(\ref{v}) taking into account all exact unperturbed orbits were proposed in the works by \citet{Holdren1970}, \citet{Chen1985}. In particular, \citet{Chen1985} considered tearing instability with a non-Maxwellian distribution function. It is also worth noting that PIC simulations are in good agreement with such methods (e.g., \citealt{Daughton2003}). For numerical procedures, the integro-differential equation is converted into a matrix equation, which is then solved to obtain the dispersion relation and the eigenfunctions. In this section, we follow the paper by \citet{Burkhart1989}, generalising their formulas to the relativistic dispersion law and the power-law distribution function.

Due to the fact that particle motion is periodic along the $x$-axis, we can rewrite the orbit integral in the form \citep{Chen1985}
\begin{equation}
\begin{split}
    S_\text{orb}=\!\int\limits_{-\infty}^t\!\!\! v_z(t')&\delta A_z(t')\text{d}t'\!=\text{i}\!\sum_{n=0}^{+\infty}\frac{\exp\left[\text{i}\left(-\omega t-n\Omega t+k_y y\right)\right]}{\omega+n\Omega-k_y v_y+\text{i}0}\times \\
    &\times \frac{1}{T}\oint \text{d}x' \frac{v_z(x')\delta A_{1z}(x')}{|v_x(x')|}\exp\left(\text{i}n\Omega t'\right),
\end{split}
\end{equation}
where $\Omega=2\pi/T$ and $T=T(\mathcal{E},P_z)$ is the period of particle motion in the $x$-direction. The $x'$ integration is carried out over one cycle of the particle motion. The singularity of the denominator shows the resonance between a particle and the electric field $\delta E_z$ of the tearing mode. Further, we are interested only in the fundamental harmonic $n=0$. Higher resonances can be discarded if the frequency of the fields is much less than the frequency of orbital motion, i.e. $\omega\ll\Omega$ for all $\mathcal{E}$ and $P_z$ \citep{Burkhart1989}. Therefore, we have
\begin{equation}\label{orb1}
\begin{split}
    S_\text{orb}=\frac{\text{i}\exp\left(-\text{i}\omega t+\text{i}k_y y\right)}{\omega-k_y v_y+\text{i}0}\frac{1}{T}\oint\text{d}x'\frac{v_z(x')\delta A_{1z}(x')}{|v_x(x')|}.
\end{split}
\end{equation}
One can easily check that approximations $v_z\approx\text{const}$ and $\delta A_{1z}\approx \text{const}$ lead to the well-known expression for the orbit integral with straight orbits (e.g. \citealt{Zelenyi1979}; see also Appendix B).

Substituting~(\ref{orb1}) into~(\ref{v}), we obtain the following integro-differential equation
\begin{equation}\label{integr}
    \frac{\text{d}^2\delta A_{1z}}{\text{d}x^2}+\Big[-k_y^2+\frac{2}{L^2\cosh^2(x/L)}\Big]\delta A_{1z}\!=\!-\frac{4\pi}{c}\!\sum_s\!\int\!\!\text{d}x' K_s(x,x')\delta A_{1z}(x'),
\end{equation}
where the integral kernel $K_s(x,x')$ is given by (see Appendix~\ref{appC})
\begin{equation}\label{kern1}
    K_s(x,x')=-2\text{i}\pi \frac{q_s^2\omega}{|k_y|c^3}\int\frac{\text{d}\mathcal{E}\text{d}P_z}{T(\mathcal{E},P_z)}\mathcal{E}\frac{\partial f_{0s}}{\partial \mathcal{E}}\frac{p_z(x)p_z(x')}{|p_x(x)||p_x(x')|},
\end{equation}
and integration is carried out over the integrals of motion, and $P_z=p_z+(q/c)A_z$ is $z$-component of a canonical momentum of a particle with electrical charge $q$.

The equation~(\ref{integr}) for $\delta A_z$ is not "local". The reason is that the amplitude of the meandering motion is large for high-energy particles, so that they "see" the perturbed field over a large range of $x$. All information about these orbits is hidden in the kernel.

The $x$ and $z$ components of kinetic momentum can be expressed in terms of the integrals of motion
\begin{equation}\label{vel}
    \begin{split}
        &p_z(x)=P_z-\frac{q_s}{c}A_z(x), \\
        &|p_x(x)|\approx\sqrt{(\mathcal{E}/c)^2-[P_z-(q_s/c)A_z(x)]^2}.
    \end{split}
\end{equation}
The limits of integration are determined by the condition $p_x^2(x)\geq 0$. Due to the fact that we use the power-law distribution function~(\ref{pwl}) without any coordinate dependence, we are able to consider the situation $r_{g0}\ll L$ only when the size of the non-magnetised region is small.

\subsection{Numerical procedure}

Let us expand the potential $\delta A_{1z}$ in the system of basis functions 
\begin{equation}
    \delta A_{1z}(x)=\sum_n \alpha_n\phi_n(x).
\end{equation}
The basis function can be chosen as pyramid functions \citep{Burkhart1989,Daughton2003}
\begin{equation}
    \phi_n(x)=
    \begin{cases}
        (x-x_{n-1})/(x_n-x_{n-1}), \quad &x_{n-1}\leq x\leq x_n, \\
        (x_{n+1}-x)/(x_{n+1}-x_{n}), \quad &x_{n}\leq x\leq x_{n+1}, \\
        0, \quad &\text{otherwise}.
    \end{cases}
\end{equation}
The basis function at the neutral plane is
\begin{equation}
    \phi_1(x)=
    \begin{cases}
        (x-x_1)/x_1, \quad &0\leq x\leq x_1, \\
        0, \quad &\text{otherwise},
    \end{cases}
\end{equation}
and the last basis function at large $x_N\gg L$ is
\begin{equation}
    \phi_N(x)=
    \begin{cases}
        (x-x_{N-1})/(x_N-x_{N-1}), \quad &x_{N-1}\leq x\leq x_N, \\
        0, \quad &\text{otherwise}.
    \end{cases}
\end{equation}
Another popular choice for basis functions is the Hermite polynomials, $H_n(\xi x)$, with the exponential weight function and $\xi=1/L$ or $\xi=k$ \citep{Daughton1999, Petri2007}. However, these functions are less appropriate for small $k$. In this case, the vector potential is practically constant throughout the entire space and varies sharply in a small region near the neutral plane. Such behaviour could hardly be fitted with the Hermite polynomials.  

One can substitute the expansion of the vector potential perturbation in~(\ref{integr}) and multiply on $\phi_m(x)$. Integrating over $x$, we obtain the matrix equation
\begin{equation}\label{system}
    \sum_n \alpha_n G_{nm}=0,
\end{equation}
where the matrix elements are
\begin{equation}\label{Gmn}
\begin{split}
    &G_{nm}(\omega,k_y)=k_y^2\!\int\!\phi_{n}(x)\phi_m(x) \text{d}x-\int\frac{\text{d}^2\phi_n}{\text{d}x^2}\phi_m\text{d}x+\\
    &+\int V_\text{ad}(x)\phi_n(x)\phi_m(x)\text{d}x+\frac{\text{i}\omega}{|k_y|c}\sum_s K_{nm},
\end{split}
\end{equation}
and
\begin{equation}
        K_{nm}=\frac{8\pi^2 q_s^2}{c^3}\int\frac{\text{d}\mathcal{E}\text{d}P_z}{T(\mathcal{E},P_z)}\mathcal{E}\frac{\partial f_{0s}}{\partial \mathcal{E}}\oint\text{d}x \frac{p_z(x)}{|p_x(x)|}\phi_m(x)\oint\text{d}x'\frac{p_z(x')}{|p_x(x')|}\phi_n(x').
\end{equation}
We changed the order of integration in $K_{mn}$; therefore, integrating $p_z(x)/|p_x(x)|$ at fixed $\mathcal{E}$ and $P_z$ should be performed only along particle trajectory, where $|p_x|\geq 0$.

Nontrivial solutions, $\alpha_n$, exist if and only if $\det \mathbf{G}(\omega,k_y)=0$. 
Following \citet{Burkhart1989}, we consider only symmetric solutions, i.e., we consider only symmetric trajectories and perform integration over $x$ and $x'$ in the first quadrant. This means that if we consider some trajectory at $x>0$, there is the mirrored one, which gives the same contribution to $K_{nm}$. However, one should be careful with "crossing" orbits: in such an approach, we can miss "half" the trajectory (which belongs to $x<0$). Therefore, when integrating "crossing" orbits in the domain $x>0$ only, it is necessary to multiply the result by 2.

One can classify all orbits on "crossing" and "non-crossing" by using integrals of motion only. In our problem, we have three integrals of motion: $y$-component of the mechanical momentum  $p_y$ (which is approximately zero for resonance particles), $z$-component of the canonical momentum $P_z$ and the energy $\mathcal{E}$. If $P_z<0$ and $\mathcal{E}^2< c^2P_z^2$, a particle has "non-crossing" orbit; at other values of $P_z$ and $\mathcal{E}$ particle has "crossing" orbits. Obviously, the contribution from electrons and positrons is the same; therefore, one can calculate the kernel only for positrons and multiply the result by 2.

\subsection{Results}

\begin{figure}
  \centering
  \includegraphics[width=10cm]{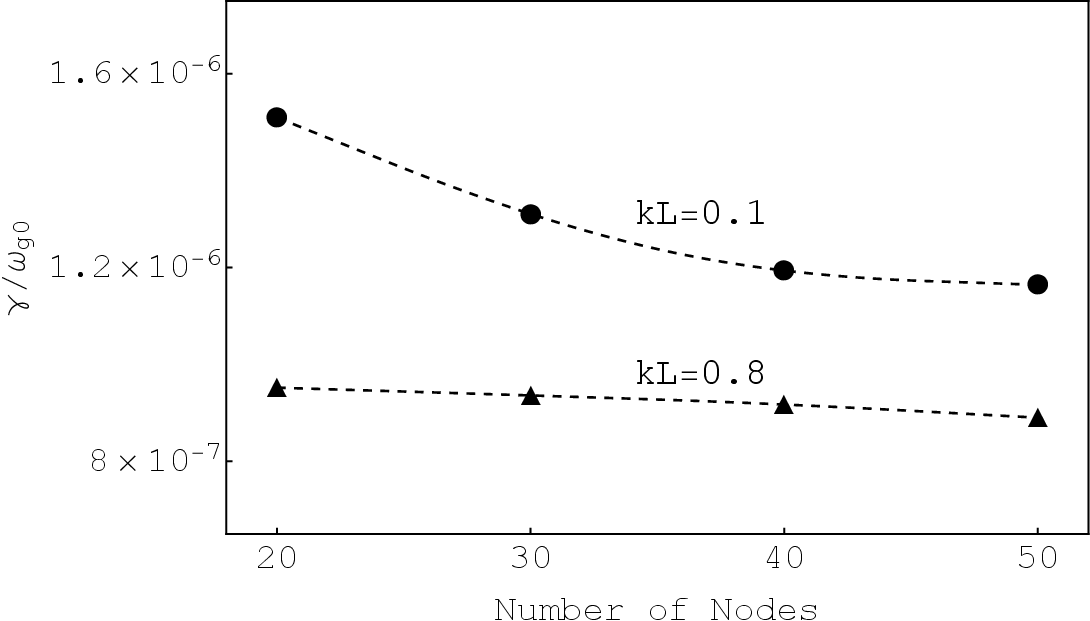}
  \caption{Convergence of the results of calculating the growth rate at $\alpha=1$ and $\epsilon=10^{-2}$. The horizontal axis indicates the number of nodes within the interval $x\in[0,3L]$}
\label{cover}
\end{figure}

The dispersion relation $\gamma(k)$ is obtained by using the exact equilibrium orbits at the parameters $r_{g0}/L=0.015$, $\mathcal{E}_\text{min}/m_e c^2=1.2$, $L=1$ cm and different $\alpha$ and $\epsilon$. According to current sheet equilibrium equations, we always have $\omega_p/\omega_{g0}=\sqrt{3/2}$, where $\omega_p=(8\pi n_0 e^2/\langle \gamma\rangle m_e)^{1/2}$ is the average plasma frequency, $\omega_{g0}=eB_0c/\langle\gamma\rangle m c^2$ is the average cyclotron frequency and $\langle\gamma\rangle =\langle\mathcal{E}\rangle/mc^2$ is the average Lorentz factor of particles. We used $N=300$ basis functions in the expansion of the vector potential perturbation. The maximum size of the computational domain is $L_\text{max}=50L$ and at $|x|>L_\text{max}$ the asymptotic solution $\sim \exp(-k|x|)$ is used. The centre points of the pyramid functions are chosen in the following way: $N_1=10$ regularly spaced points within interval $x\in[0,3l_\text{min}]$, $N_2=20$ points within interval $x\in [3l_\text{min},L]$ $N_3=10$ points for $x\in[L,3L]$ and $N_4=260$ points within interval $x\in[3L,L_\text{max}]$. Increasing the computational domain to size $x\in[0,70L]$ does not lead to changes in the results. All integrals were calculated in Wolfram Mathematica using the "Local Adaptive" method. We calculated the growth rate for 9 points in the interval $0.1\leq kL\leq 0.9$ and used the quadratic interpolation. 
One can see in Figure~\ref{f1} that our estimates provide a good approximation for the growth rate. 
Figure~\ref{cover} shows the convergence of our calculations. For this purpose, we took a different number of basis functions in the interval $x\in[0,3L]$, in which the current of resonant particles is non-zero. 

\begin{figure}
  \centering
  \includegraphics[width=11cm]{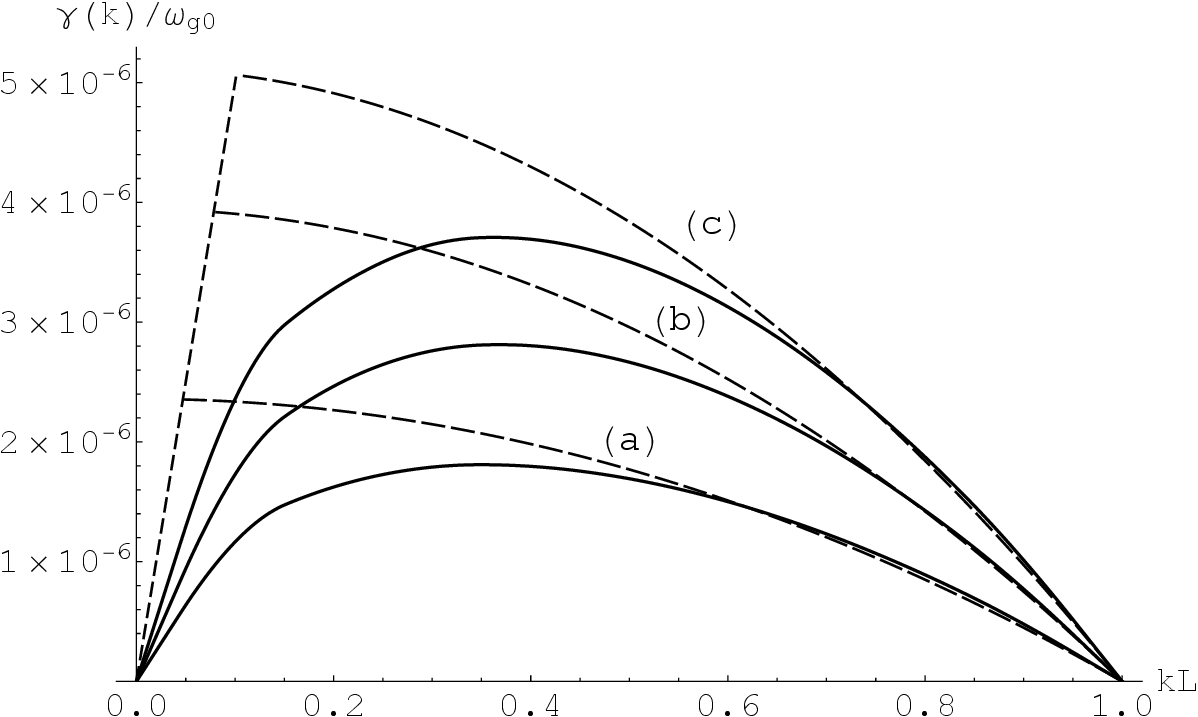}
  \caption{The dependence of the growth rate $\gamma(k)/\omega_{g0}$ on $kL$ for $r_{g0}/L=0.015$,  and (a) $\alpha=1$, $\epsilon=10^{-2}$, (b) $\alpha=2$, $\epsilon=1.5\times10^{-2}$, (c) $\alpha=3$, $\epsilon=5\times 10^{-3}$. The solid lines are the solution of the equation~(\ref{integr}); the dashed lines are the estimations~(\ref{grwth}) and~(\ref{est})}
\label{f2}
\end{figure}

The wave number $k_*$ at which the maximum growth rate is observed is always the same, and $k_*L\sim 0.4$. The same is true for the non-relativistic plasma, and this result does not depend on whether exact particle trajectories are considered or the approximation of straight trajectories.

\begin{figure}
  \centering
  \includegraphics[width=11cm]{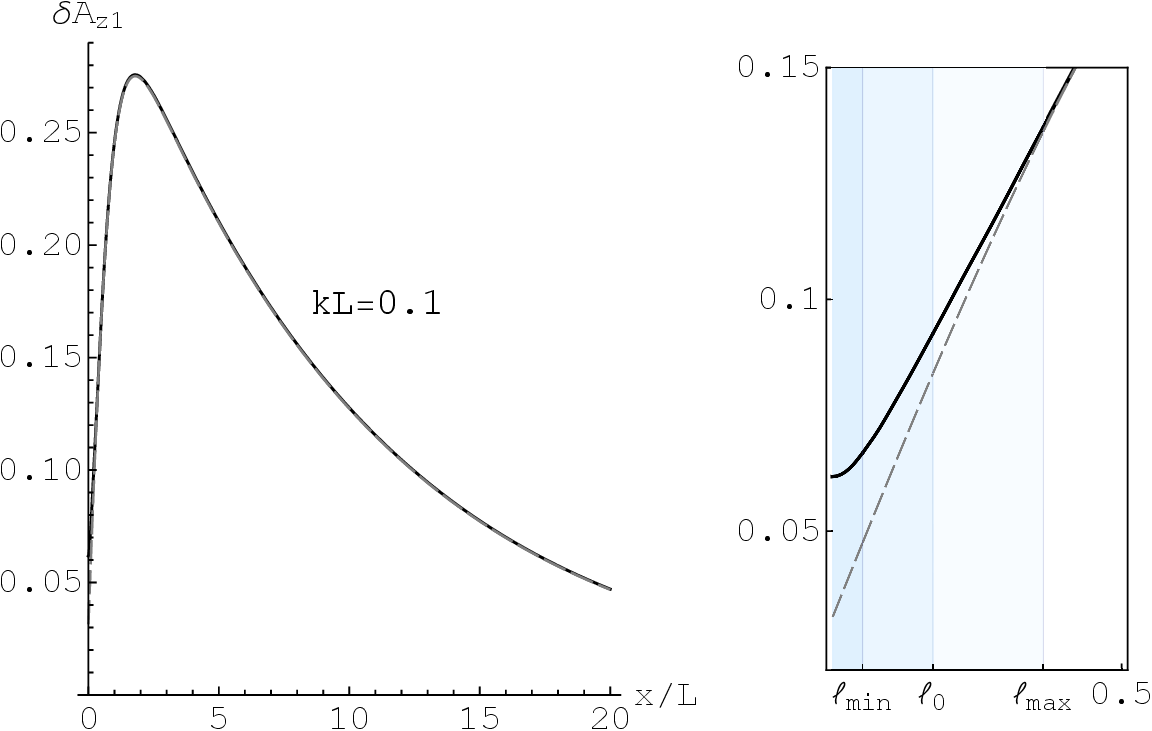}
  \caption{The normalized eigenfunction $\delta A_{z1}(x)$ at $kL=0.1$ and $r_{g0}/L=0.015$, $\alpha=1$, $\epsilon=10^{-2}$; the gray dashed line is the outer solution~(\ref{outer}); gray shaded areas shows inner regions where particles of different energies perform meandering motion}
\label{eign}
\end{figure}

Figure~\ref{eign} shows the normalized eigenfunction for $kL=0.1$. We chose this eigenfunction because it has the worst convergence in our calculations since it extends over a large distance. One can see the deviation from the outer solution near the neutral plane, where particles decouple from the fluid motion. Indeed, the largest deviation is observed at $|x|\lesssim l_\text{min}$ as we used in our theoretical considerations.

\section{Discussion and conclusions}

Let us estimate the dependence of the growth rate on the power-law index of the particle spectrum, $\alpha$. It is worth comparing only the maximum growth rates. It should be taken into account that the maximum growth rate is achieved when the layer thickness is of the order of the characteristic Larmor radius, $r_{g0}\sim L$. Even though our calculations assumed $r_{g0}\ll L$, the obtained results could be used as an order of magnitude estimate even at  $r_{g0}\sim L$. 

For $\alpha\geq 2$, there is only one energy scale $\mathcal{E}\sim\mathcal{E}_\text{min}$ as in the Maxwellian case. Therefore, $\gamma(\alpha\geq2)$ is of the order of the tearing growth rate for the Maxwellian distribution function with $k_B T\sim\mathcal{E}_\text{min}$. In this case, we could use $L\sim \mathcal{E}_\text{min}/eB_0$. At $\alpha<2$, the minimal width of the current sheet is of the order of the Larmor radius of the energetic particles, $L\sim \mathcal{E}_\text{max}/eB_0$.  Assuming that the magnetic field $B_0$ is the fixed external parameter, we obtain
\begin{equation}
    \frac{\gamma(\alpha<2)}{\gamma(\alpha> 2)}\sim \epsilon^{3/2}\ll1
\end{equation}
Therefore, the growth rate is strongly suppressed if the particle spectrum is shallow, $\alpha< 2$.


The slow linear stage of the tearing instability in a relativistic pair plasma can be explained in the same way as in the case of an ion-electron plasma. Only low-energy particles with a small Larmor radius participate in the tearing instability, while high-energy particles do not make a significant contribution. One might naively assume that applying a uniform magnetic field $\boldsymbol{B}_n$ that is perpendicular to the current sheet and $B_n/B_0\ll 1$ would magnetize low-energy particles in the inner region so that only high-energy particles would be involved in the tearing instability, thus enhancing the instability. However, as was shown by \cite{Coroniti1980} and \cite{Lembege1982}, for the ion-electron plasma, the Hall drift of low-energy magnetised electrons in crossed $\delta \boldsymbol{E}$ and $\boldsymbol{B}_n$ fields makes the current sheet inhomogeneous. Moreover, such inhomogeneities create a large electric field in order to maintain the charge neutrality with ions, which also causes the ions to drift, so the entire plasma is subject to compression.
If the energy required for such compression exceeds the released free energy due to the pinching of current filaments, the ion tearing instability is suppressed. However, 3D simulations show that the collisionless reconnection instability might be possible even in the presence of the normal magnetic field $\boldsymbol{B}_n$ (\citealt{Buchner1996}, \citealt{Buchner1999}), contrary to what is observed in 2D simulations. 
Also, in the relativistic pair plasma, non-magnetised high-energy particles that we can think of as "ions" due to large relativistic mass do not feel any additional electric field because low-energy particles already have zero net electric charge. Therefore, these high-energy particles move independently from the magnetised low-energy particles, and this requires further consideration.


Similarly, it is necessary to consider the tearing instability with a guide field parallel to the current sheet. Firstly,  without the guide field, the three-dimensional current sheet in the pair plasma is unstable with respect to the drift-kink instability \citep{Zenitani2005}, and the growth rate of this instability is several times greater than that of the tearing mode.
Secondly, the tearing instability with the guide field is of significant interest for the theory of turbulence in the relativistic pair plasma. As was shown for the non-relativistic case, the MHD turbulent cascade produces small-scale current sheets that could be destroyed by the reconnection process (e.g. \citealt{Mallet2017,Loureiro2017}). 
A similar process may occur in the relativistic plasma, which can lead to efficient acceleration of particles.
Therefore, the case of the tearing instability with the guide field deserves special attention and will be considered in subsequent publications.


Let us briefly summarise the results obtained. In this work, we considered the collisionless tearing instability in a relativistic pair plasma with a power-law distribution function. An analytical expression is obtained for the instability growth rate. The analytical results are compared with the numerical solution that takes into account all unperturbed exact particle trajectories. As a result, we found that the tearing instability is suppressed at the harder spectrum. 



\section*{Funding}

This research was supported by the Israel Science Foundation under the grant 2067/19.





\appendix

\section{Derivation of the vector potential equation}\label{appA}

In considered geometry with the given magnetic field profile $f(x)=\tanh(x/L)$, the equilibrium vector potential $A_z$ depends only on the $x$-coordinate and has the form
\begin{equation}
A_z(x)=-B_0L\ln\cosh(x/L),
\end{equation}
where boundary conditions $\text{d}A_z/\text{d}x=0$ at $x=0$ and $A_z\rightarrow \pm B_0x$ at $x\rightarrow\pm\infty$ are assumed.

Therefore, there are three integrals of motion for each charged particle
\begin{equation}\label{2-01}
p_y,\quad P_z=p_z+\frac{q}{c}A_z, \quad \mathcal{E}=\sqrt{p^2c^2+m^2c^4},
\end{equation}
where $p_y$ is $y$-component of the mechanical momentum of a particle, $P_z$ is $z$-component of a canonical momentum of a particle with electrical charge $q$, and $\mathcal{E}$ is its mechanical energy. Further, particles are assumed to be ultrarelativistic with the dispersion law $\mathcal{E}=pc$.

The equilibrium distribution function $f_{0s}$ (where $s$ denotes the sort of particles) has to depend only on these integrals of motion since, in this case, the Vlasov equation is satisfied identically. 
There are infinitely many such distribution functions, and all of them satisfy the current sheet equilibrium equation $P_\text{tot}(x)+B^2(x)/8\pi=\text{const}$, where plasma pressure is written as
\begin{equation}
    P_\text{tot}=\frac{1}{3}\sum_s\int\frac{p^2 c^2}{\mathcal{E}}f_{0s}\text{d}\boldsymbol{p}.
\end{equation}
Now, we do not determine the explicit form of $f_{0s}$. We only assume that it depends on the integrals of motion. Only one condition on the distribution function is imposed: within the thin layer near $x=0$, the dependence on canonical momentum $P_z$ disappears, and it has the following form in the momentum space:
\begin{equation}\label{dist_func}
\begin{cases}
f_{0s}=C_m, \quad &\mathcal{E}<\mathcal{E}_\text{min} \\
    f_{0s}=C_p \mathcal{E}^{-(\alpha+2)}\exp\left(-\mathcal{E}/\mathcal{E}_\text{max}\right), \quad &\mathcal{E}\geq\mathcal{E}_\text{min},
\end{cases}
\end{equation}
where constants $C_m$ and $C_p$ can be found from the condition of continuity of the distribution function at $\mathcal{E}=\mathcal{E}_\text{min}$ and from the normalization condition:
\begin{equation}\label{constants}
    C_m=\frac{n_0 c^3}{4\pi (\mathcal{E}_\text{min})^3}\left[\text{E}_\alpha(\epsilon)+1/3\right]^{-1}, \quad C_p=\frac{n_0 c^3}{4\pi }\mathcal{E}_\text{min}^{\alpha-1}\left[\text{E}_\alpha(\epsilon)+1/3\right]^{-1}.
\end{equation}
where $\text{E}_\alpha(\epsilon)$ is the exponential integral function
\begin{equation}
\text{E}_\alpha(\epsilon)=\int\limits_1^{+\infty}\frac{\exp(-\epsilon x)}{x^\alpha}\text{d}x.
\end{equation}
The macroscopic drift velocity of plasma is not assumed to be constant;  it is found from
\begin{equation}\label{drift}
    U_s(x)=\frac{c}{8\pi q_s n(x)}\frac{\text{d}B}{\text{d}x}.
\end{equation}
Actually, this drift is caused by both the magnetization current $\mathbf{j}_m=-c\,\text{rot}(P\mathbf{B}/B^2)$ that is induced by a non-uniform distribution of Larmor circles across the current sheet and the $\nabla B$-current $\mathbf{j}_b=cP[\mathbf{B}\times \nabla B]/B^3$ (e.g. \citealt[pp. 91-93]{Bellan2006}). The total current density is $\mathbf{j}=\mathbf{j}_m+\mathbf{j}_b=-c [\nabla P\times\mathbf{B}]/B^2=(c/4\pi)(\text{d}B/\text{d}x)\hat{\mathbf{z}}$ that leads to expression~(\ref{drift}) for the drift velocity. The same result can be obtained from the MHD equation $(1/c)[\,\mathbf{j}\times\mathbf{B}]=\nabla P$.

Since the reversal magnetic field profile satisfy to the condition $B\sim B_0x/L$ at $x\rightarrow 0$, at $x=0$ we have
\begin{equation}\label{U0}
    \frac{U_0}{c}=\frac{1}{8\pi}\frac{B_0}{en_0 L}=\frac{2\langle\mathcal{E}\rangle}{3eB_0 L}
\end{equation}
We assume that $U_0\ll c$. It means that average the Larmor radius $r_{g0}\sim\langle \mathcal{E}\rangle/eB_0$ is much smaller than the current sheet thickness $L$. Thus, the theory is developed only for thick current sheets. When the drift velocity may be close to the speed of light, the tearing instability is stabilized \citep{Hoshino2020}.

We investigate only low-frequency tearing oscillations when $\omega \ll ck$. This means that the phase velocity of perturbations is much less than the speed of light. In this case, we can neglect the displacement current and perturbations of the scalar potential $\delta\phi$. Due to the development of the tearing instability in the system, perturbation of the vector potential $\delta \boldsymbol{A}$ appears, for which the Maxwell equation has the form
\begin{equation}\label{ampere}
    \nabla^2 \delta \boldsymbol{A}=-\frac{4\pi}{c}\delta\boldsymbol{j},
\end{equation}
where current perturbation $\delta\boldsymbol{j}$ is determined by the perturbation of the distribution function $\delta f_s$.  The solution of the Vlasov equation is sought in the form $f_s=f_{0s}+\delta f_s$, where $f_{0s}=f_{0s}(\mathcal{E},P_z)$ is the equilibrium distribution function. Therefore, the linearized Vlasov equation is
\begin{equation}\label{6-01}
\frac{\text{d}\delta f_s}{\text{d} t}=-q_s\left(\delta\boldsymbol{E}+\frac{1}{c}\boldsymbol{v}\times\delta\boldsymbol{B}\right)\cdot\frac{\partial f_{0s}}{\partial\boldsymbol{p}}
\end{equation}
Perturbations of electric and magnetic fields can be expressed in terms of vector potential
\begin{equation}\label{6-02}
\delta\boldsymbol{E}=-\frac{1}{c}\frac{\partial\delta\boldsymbol{A}}{\partial t}, \quad \delta\boldsymbol{B}=\nabla\times\delta\boldsymbol{A}
\end{equation}
Further, it is assumed that the time and coordinate dependence of all perturbed quantities has the form
\begin{equation}\label{6-03}
\delta\psi=\delta\psi_1(x)\exp\left(-\text{i}\omega t+\text{i}\boldsymbol{k}\cdot\boldsymbol{r}\right),
\end{equation}
where $\boldsymbol{k}=k\hat{\boldsymbol{y}}$ (it corresponds to the most unstable oscillations). It also means that $k_z=0$ and $\boldsymbol{k}\cdot\boldsymbol{U}_s=0$, 
According to~(\ref{6-03}), one can write $\delta\boldsymbol{E}=\text{i}(\omega/c)\delta\boldsymbol{A}$ and $\delta\boldsymbol{B}=\text{i}\boldsymbol{k}\times\delta\boldsymbol{A}-(\partial\delta A_z/\partial x)\hat{\boldsymbol{y}}$.

Performing time integration of the equation~(\ref{6-01}), and assuming that $\delta f_s(t=-\infty)=0$, we obtain
\begin{equation}\label{6-05}
\delta f_s=-\frac{q_s}{c}\left[-\frac{\partial f_{0s}}{\partial P_z}\delta A_z+\text{i}\omega \frac{\partial f_{0s}}{\partial \mathcal{E}}\!\int\limits_{-\infty}^{t}(\boldsymbol{v}\cdot\delta\boldsymbol{A})\text{d}t'\right].
\end{equation}
It should be noted that the particle's coordinates and velocities under the integral are determined by solving the equations of motions for an electron/positron in the unperturbed electromagnetic field. In this case, quantities $\mathcal{E}$ and $P_z$ are still integrals of motion and they are conserved along any particle trajectory, therefore, we can take $\partial f_{0s}/\partial \mathcal{E}$ and $\partial f_{0s}/\partial P_z$  out from under the integral. Also, we used the identity $(\omega-\boldsymbol{k}\cdot\boldsymbol{v})\delta\psi+\text{i}v_x\partial\psi/\partial x=\text{i}\text{d}(\delta\psi)/\text{d}t$.

The current perturbation is
\begin{equation}\label{6-06}
\delta \boldsymbol{j}=\sum_s q_s\!\int\boldsymbol{v}\delta f_s\text{d}\boldsymbol{p}.
\end{equation}
In our case, we obtain that only the $z$-component of the current perturbation is not zero. The current perturbation can be represented as a sum of two terms \citep[p. 312]{Galeev1984}
\begin{equation}\label{6-07}
\begin{split}
\delta j_z=-\sum_s\frac{q^2_s}{c}\Big[-\delta A_z\int\frac{\partial f_{0s}}{\partial P_z}v_z\text{d}\boldsymbol{p}\,+\text{i}\omega\int \frac{\partial f_{0s}}{\partial \mathcal{E}}v_z\int\limits_{-\infty}^t v_z(t')\delta A_z(t')\text{d}t' \text{d}\boldsymbol{p}\Big].
\end{split}
\end{equation}
The first one, $\delta \boldsymbol{j}^{\text{ad}}$, is the adiabatic perturbation of the current, which arises due to the slow change in the magnetic field topology (the magnetised motion of particles resulting from the slow plasma convection motion). The second one, $\delta \boldsymbol{j}^{\text{res}}$, is the resonant current density perturbation due to the resonant interaction of particles with wave perturbations.

Substituting this current perturbation into~(\ref{ampere}), we obtain
\begin{equation}\label{vect}
\frac{\text{d}^2}{\text{d}x^2}\delta A_z+\left[-k_y^2-V_\text{ad}(x)\right]\delta A_z=4\pi\text{i}\omega \sum_s\frac{q_s^2}{c^2}\int \frac{\partial f_{0s}}{\partial \mathcal{E}}v_z\int\limits_{-\infty}^t v_z(t')\delta A_z(t')\text{d}t' \text{d}\boldsymbol{p}.
\end{equation}
Other components of $\delta \boldsymbol{A}$ vanish in our geometry. The adiabatic potential $V_\text{ad}(x)$ characterises the adiabatic perturbation of the current density and is determined as follows:
\begin{equation}\label{6-09}
V_\text{ad}(x)=4\pi\sum_s\frac{q_s^2}{c^2}\int \frac{\partial f_{0s}}{\partial P_z}v_z \text{d}\boldsymbol{p}
\end{equation}
Rewriting $\partial f_{0s}/\partial P_z=(c/q_s)\partial f_{0s}/\partial A_z$ and taking the derivative $\partial/\partial A_z$ out of the integral, we obtain $V_\text{ad}(x)=(4\pi/c)\partial j_z/\partial A_z$. It also means that $\delta j_\text{ad}=V_\text{ad}(x)\delta A_z$ is just the second term in the Taylor series for the current density $j_z(A_z+\delta A_z)$. Since $\partial j_z/\partial A_z=(\text{d} j_z/\text{d} x)(\text{d} A_z/\text{d} x)^{-1}$, one can see that $V_\text{ad}(x)=B''_y(x)/B_y(x)$. Therefore, this potential is determined by the equilibrium: if we know the profile of the magnetic field, we can easily find the adiabatic potential. The same result is obtained from the MHD equations \citep{Sturrock1994}. It should be noted that for this conclusion, we used the fact that the distribution function depends only on integrals of motion.

Assuming $B(x)=B_0\tanh(x/L)$, we obtain the well-known formulae
\begin{equation}\label{Vadd}
    V_\text{ad}(x)=-\frac{2}{L^2}\frac{1}{\cosh^2(x/L)}.
\end{equation}

\section{The integral kernel}\label{appC}
The kernel $K_s(x,x')$ is given by
\begin{equation}\label{kernC}
    K_s(x,x')=\frac{q_s^2\omega}{c}\!\!\!\int\!\!\frac{\text{d}\boldsymbol{p}}{T}\frac{\partial f_{0s}/\partial \mathcal{E}}{(\omega-k_y v_y+\text{i}0)}\frac{v_z(x)v_z(x')}{|v_x(x')|}.
\end{equation}
This integral is symmetric in momentum $p_x$; therefore, one can multiply the integral by the factor of two and integrate over $p_x$ from $0$ to $+\infty$. Next, when we decided on the sign of $p_x$, let us to rewrite differential $\text{d}\boldsymbol{p}$ in terms of integrals of motion as $(\mathcal{E}/c^2|p_x|)\text{d}\mathcal{E}\text{d}p_y\text{d}P_z$. 
Also, the main contribution in~(\ref{kernC}) is given by the semi-residue $\omega=k_y v_y$. Therefore, we have
\begin{equation}
    K_s(x,x')=-2\text{i}\pi \frac{q_s^2\omega}{c}\int\frac{\text{d}\mathcal{E}\text{d}p_y\text{d}P_z}{T(\mathcal{E},P_z)}\frac{\partial f_{0s}}{\partial \mathcal{E}}\delta\!\left(\omega-\frac{k_y p_y c^2}{\mathcal{E}}\right)\frac{p_z(x)p_z(x')}{|p_x(x)||p_x(x')|}.
\end{equation}
We are interested in the first order in $\omega$ since it is a small quantity ($\omega \ll k_y c$), therefore one can put $p_y=(\mathcal{E}/c)(\omega/k_yc)\approx 0$ under the integral. As a result, we obtain
\begin{equation}
    K_s(x,x')=-2\text{i}\pi \frac{q_s^2\omega}{|k_y|c^3}\int\frac{\text{d}\mathcal{E}\text{d}P_z}{T(\mathcal{E},P_z)}\mathcal{E}\frac{\partial f_{0s}}{\partial \mathcal{E}}\frac{p_z(x)p_z(x')}{|p_x(x)||p_x(x')|}.
\end{equation}
Now one can also come to the approximation of straight orbits if we assume that $p_z$, $p_x\approx \sqrt{\mathcal{E}^2/c^2-P_z^2}$ and $\delta A_z$ do not depend on spatial coordinates within the inner region. The integration over the canonical momentum $P_z$ leads to
\begin{equation}
    \frac{4\pi}{c}\sum_s \int\text{d}x' K_s(x,x')\delta A_z(x')\rightarrow \left(\frac{8\pi^3\text{i}\omega e^2}{kc^4}\int\frac{\partial f_{0s}}{\partial \mathcal{E}}\mathcal{E}^2\text{d}\mathcal{E}\right)\delta A_z\equiv V_\text{res}\delta A_z.
\end{equation}
In this case, the equation~(\ref{vect}) is reduced to the "Schrödinger-type" equation
\begin{equation}
\frac{\text{d}^2}{\text{d}x^2}\delta A_z+\left[-k_y^2-V_\text{ad}(x)-V_\text{res}(x)\right]\delta A_z=0.
\end{equation}
For this reason, the function $V_\text{res}(x)$ is sometimes called the resonant potential. It represents a narrow potential barrier with a width $\sim 2l_\text{min}$, and at larger distances, it vanishes. This equation is similar to that used by \cite{Zelenyi1979}. Using their procedure, one can find the growth rate of the tearing instability $\gamma(k)$. In this case, the same expression as~(\ref{grwth}) is obtained, but with a numerical coefficient $2\sqrt{2}/\pi\sim 0.9$ instead of $0.5$. 

There are two more physical reasons to believe that the numerical value of the growth rate is smaller than these estimates give. The first reason is that the contribution from trajectories only within the interval $|x|\lesssim l_\text{min}$ is taken into account; in fact, this interval is wider (see Fig.~\ref{eign}). We also did not take into account the contribution from non-crossing orbits, whose contribution is comparable to that from crossing orbits for $|x|\sim l_\text{min}$. This additionally increases the plasma conductivity near the neutral layer and reduces the growth rate.

Let us return to the matrix equations~(\ref{system}) and~(\ref{Gmn}), where the quantity $K_{mn}$ was introduced:
\begin{equation}
    \frac{\text{i}\omega}{|k_y|c}\sum_s K_{mn}=-\frac{4\pi}{c}\sum_s\int\text{d}x\int\text{d}x'\phi_m(x)K_s(x,x')\phi_n(x')
\end{equation}
For the numerical calculation, it is convenient to use dimensionless variables
\begin{equation}
    \hat{p}=\frac{P_z c}{\langle \mathcal{E}\rangle}, \quad \hat{\mathcal{E}}=\frac{\mathcal{E}}{\langle \mathcal{E}\rangle}, \quad \hat{T}=\omega_{g0} T, \quad \hat{x}=\frac{x}{L}, \quad \hat{k}=k_y L 
\end{equation}
where $\langle \mathcal{E} \rangle$ is the average particle energy which is determined by
\begin{equation}\label{en}
\langle\mathcal{E}\rangle=\frac{3\left[\text{E}_{\alpha-1}(\epsilon)+1/4\right]}{3\text{E}_{\alpha}(\epsilon)+1}\mathcal{E}_\text{min},
\end{equation}
and $\omega_{g0}=|q_s| B_0c/\langle \mathcal{E}\rangle$ is the average cyclotron frequency in the asymptotic magnetic field. Taking into account that we consider ultarelativistic particles, for the Harris current sheet we have
\begin{equation}
    \hat{p}_z=\hat{p}+\frac{L}{r_{g0}}\ln\cosh\hat{x}, \quad |\hat{p}_x|\approx\left[\hat{\mathcal{E}}^2-\left(\hat{p}+\frac{L}{r_{g0}}\ln\cosh\hat{x}\right)^2\right]^{1/2}
\end{equation}
Substituting the distribution function~(\ref{dist_func}), we obtain
\begin{equation}
    K_{nm}=\frac{9}{8L}\left(\frac{L}{r_{g0}}\right)^3\frac{\left(3\text{E}_\alpha(\epsilon)+1\right)^{\alpha-2}}{[3\text{E}_{\alpha-1}(\epsilon)+3/4]^{\alpha-1}}\hat{K}_{nm}
\end{equation}
where dimensionless matrix elements are
\begin{equation}
\begin{split}
&\hat{K}_{nm}=\!\int\!\!\frac{\text{d}\hat{\mathcal{E}}\text{d}\hat{p}}{\hat{T}(\hat{\mathcal{E}},\hat{p})}\hat{\mathcal{E}}^{-(\alpha+2)}\exp\left(-\frac{\langle \mathcal{E}\rangle}{\mathcal{E}_\text{max}}\hat{\mathcal{E}}\right)\times \\
    &\left(\alpha+2+\frac{\langle \mathcal{E}\rangle}{\mathcal{E}_\text{max}}\hat{\mathcal{E}}\right)\! \oint\!\text{d}\hat{x} \frac{\hat{p}_z(x)}{|\hat{p}_x(x)|}\hat\phi_m(x)\oint\text{d}\hat{x}'\frac{\hat{p}_z(x')}{|\hat{p}_x(x')|}\hat\phi_n(x')
\end{split}
\end{equation}


\bibliographystyle{jpp}

\bibliography{References}

\end{document}